\documentclass[letter]{aa}
\usepackage{cite}
\usepackage{graphicx}
\usepackage{txfonts}
\usepackage{natbib}
\bibpunct{(}{)}{;}{a}{}{,} 

\usepackage{longtable}

\begin{document}
	\title{Searching for galactic sources in the \emph{Swift} GRB catalog}

	\subtitle{Statistical analyses of the angular distributions of FREDs}

	\author{J.C. Tello
		\inst{1}
		\and
		A.J. Castro-Tirado\inst{1}
		\and
		J. Gorosabel\inst{1}
		\and
		D. P\'erez-Ram\'irez\inst{2}
		\and
		S. Guziy\inst{3}
		\and
		R. S\'anchez-Ram\'irez\inst{1}
		\and
		M. Jel\'inek\inst{1}
		\and
		P. Veres \inst{4,5}
		\and
		Z. Bagoly \inst{4}
	}
	\institute{Instituto de Astrof\'isica de Andaluc\'ia (I.A.A.-C.S.I.C.), Spain
		\and
		Universidad de Ja\'en, Spain
		\and
		Nikolaev National University, Ukraine
		\and
		E\"otv\"os University, Budapest
		\and
		Bolyai Military University, Budapest
		}
		
		\abstract
		{Since the early 1990s, gamma ray bursts (GRB) have been accepted to be of extra-Galactic origin because of the isotropic distribution observed by BATSE and the redshifts observed via absorption line spectroscopy. Nevertheless, upon closer examination at least one case turned out to be of Galactic origin. This particular event presented a fast rise, exponential decay (FRED) structure, which leads us to believe that other FRED sources might also be Galactic.}
		{This study was set out to estimate the most probable degree of contamination by Galactic sources that certain samples of FREDs have.}
		{To quantify the degree of anisotropy, the average dipolar and quadripolar moments of each sample of GRBs with respect to the Galactic plane were calculated. This was then compared to the probability distribution of simulated samples comprising a combination of isotropically generated sources and Galactic sources.}
		{We observe that the dipolar and quadripolar moments of the selected subsamples of FREDs are found more than two standard deviations outside those of random isotropically generated samples. The most probable degree of contamination by Galactic sources for the FRED GRBs of the \emph{Swift} catalog detected until February 2011 that do not have a known redshift is about 21 out of 77 sources, which represents roughly 27\%. Furthermore, we observe that by removing from this sample those bursts that have any type of indirect redshift indicator and multiple peaks, the most probable contamination increases to 34\% (17 out of 49 sources).}
		{It is probable that a high degree of contamination by Galactic sources occurs among the single-peak FREDs observed by \emph{Swift}. Accordingly we encourage additional studies on these type of events to determine the nature of what could be an exotic type of Galactic source.}
		
		\maketitle

\section{Introduction}
Gamma ray bursts (GRBs) are the most energetic explosions known in the Universe, second only to the Big Bang. Discovered in the 1960s, they were widely believed to originate in the Milky Way because of their relatively high flux of photons, which needs an unprecedented emission mechanism to account for this high energy output. It was not until 1997 when the first measurement of redshift was performed on a GRB afterglow that the cosmological nature of these objects was asserted without doubt \citep{1997Natur.387..878M}.

The GRB afterglows fade within a few hours, and as a consequence, the redshift of most GRBs are unknown. In the past several studies have been carried out to indirectly determine the Galactic or extra-Galactic nature of the bursts by analyzing their spatial distribution in the sky \citep{1981Ap&SS..80....3M, 1992Natur.355..143M, 1998A&A...339....1B}, and historically it served as a strong argument against the Galactic origin of GRBs \citep{1999ApJS..122..465P}. This technique has also been used to suggest a more local nature of long-lag bursts by showing that they may be related to the super-Galactic structure \citep{2002ApJ...579..386N, 2008A&A...484..143F}.

The observed light curve of each GRB varies from burst to burst, particularly during in the prompt phase when the gamma ray emission is emitted, where one or multiple peaks with a variety of shapes are observed. However, some of them present a fast rise and exponential decay (FRED hereafter) behavior. These have been correlated with other properties of the bursts \citep{1994ApJ...426..604B}, suggesting that they may be of a different nature than other GRBs. 

There has been at least one reported GRB that upon closer examination has resulted to be a phenomenon from within the Milky Way \citep{2008Natur.455..506C,2008Natur.455..503S}. This source displayed a FRED structure, which leads us to believe that there could be others like it.

We aim to  estimate the most probable degree of contamination by Galactic sources that certain samples of FREDs have. We have organized the paper as follows: in Section~2 we establish the selection criteria of the studied samples. Section~3 describes the methodology used for quantifying the anisotropy and determining the probability of observing these values for both extra-Galactic and Galactic sources while taking into account the exposure of \emph{Swift}. We discuss the results from our analysis in Section~4 and give our main conclusions in Section~5.

\section{Sample selection}
To achieve a homogeneous distribution, only \emph{Swift}-detected GRBs were taken into account. From the catalog of 596 GRBs detected by \emph{Swift} before March 2011, 111 GRBs were selected because they had a FRED structure reported in a GCN. Using the information available in peer-reviewed papers\footnote{Only two peer-reviewed papers were relevant for the sample selection \citep{2011AA...529A.110C,2009AJ....138.1690P}} and other GCN circulars related to the 111 FRED GRBs, the following subsamples were selected:
\footnote{For a list of specific selected bursts see Appendix A} 
	\begin{itemize}
	\item \textbf{Sample 1:} All 111 FREDs detected by \textit{Swift} until February 2011.
	\item \textbf{Sample 2:} 77 FREDs without any measured redshift.
	\item  \textbf{Sample 3:} 71 FREDs without stated high-redshift criteria.
	\item  \textbf{Sample 4:} 59 FREDs without any type of indirect redshift indication.
	\item  \textbf{Sample 5:} 49 FREDs without any redshift indications or multiple peaks.
	\end{itemize}

\begin{figure}[htb]
	\centering
	\includegraphics[width=0.4\textwidth]{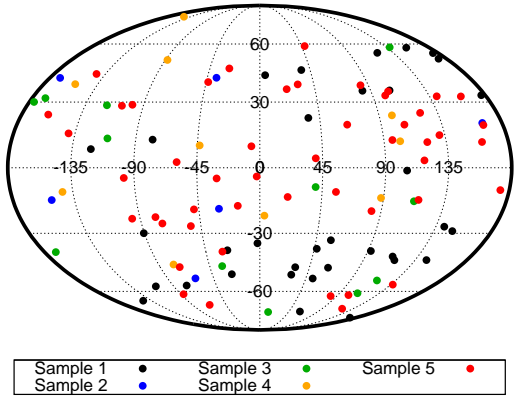}
	\caption{Sources in each one of the samples.To avoid redundancy, only the sources not present in the subsequent samples were included.}
	\label{Samples}
\end{figure}

It is important to note that only sample 5 included solely those bursts that consisted of one pure FRED peak.

\section{Anisotropy quantification}	
It has been proven \citep{1989ApJ...346..960H} that the mean dipolar and quadripolar moments of the Galactic coordinates ($cos b$ and $sin^2 b $, where $b$ is the Galactic latitude) are good tools to quantify the isotropy with respect to the Galactic plane~\citep{1994wedo.book.....C}. The degree of isotropy of each sample was calculated using the coordinates available from the gamma-ray burst coordinate network (GCN) circulars for each burst. The results are shown in Table~\ref{SampleMoments}.

\begin{table}
\centering
	\begin{tabular} {| c | c | c |}
		\hline
		Sample & $ <cos b> $ & $<sin^2 b>$ \\ \hline
		\#1 & 0.7883 & 0.3397 \\ \hline
		\#2 & 0.8221 & 0.2860 \\ \hline
		\#3 & 0.8184 & 0.2909 \\ \hline
		\#4 & 0.8344 & 0.2673 \\ \hline
		\#5 & 0.8397 & 0.2622 \\
		\hline
	\end{tabular}
	\caption{Dipolar and quadripolar moments of the samples as a quantitative measurement of the degree of isotropy in the samples.}
	\label{SampleMoments}
\end{table}

\subsection{Exposure map}
Owing to the nature of its instruments, orbit, and mission, \textit{Swift}'s pointing toward the sky is not homogeneous. It is of particular relevance to note that there has been less integrated exposure time toward the Milky Way's disk than toward the Galactic poles. This fact would represent a bias for the nature of the study carried out for this publication if left unaccounted, therefore we created a map by integrating the exposure mask function for the BAT instrument, multiplied by the exposure times of all observations carried out between April 16, 2005 and February 1, 2011, taking into account the pointing and rotation of the BAT instrument\footnote{The method used to derive the exposure map is the same as the one detailed in Veres \emph{et al.}, 2010\nocite{2010AIPC.1279..457V}}.
\begin{figure}[htb]
	\centering
	\includegraphics[width=0.4\textwidth]{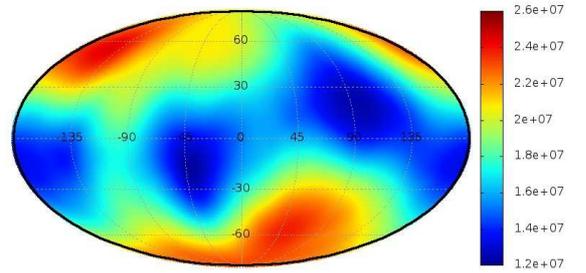}
	\caption{Swift exposure map in Galactic coordinates derived for this study. Colors represent the exposure time (in seconds).}
	\label{Exposure}
\end{figure}

\subsection{Monte Carlo simulations}
Monte Carlo simulations were carried out to determine the probability mass function (PMF) of the average dipolar and quadripolar moments of random GRB distributions. Therefore random coordinates were generated, taking care that they had a homogeneous distribution on an spherical surface.

These random points were then used to determine the PMF of the dipolar and quadripolar moments of random sources in the sky to determine by how much the observed samples' values deviated from those of a completely isotropically generated sample. To do this we generated an equal number of random points to that of each sample, recording the value of the mean dipolar and quadripolar moment and iterating a statistically significant number of times $10^6-10^9$ iterations. The histogram of the recorded values was then used to determine the values for standard deviations ($\sigma, 2\sigma, 3\sigma$).

\subsection{Metropolis-Hastings algorithm}
To account for the anisotropy of \textit{Swift}'s exposure of the night sky, it was necessary to factor in the probability that a particular random source was detected by \textit{Swift}. We used the Metropolis-Hastings algorithm for this, which can be summarized in three steps:

\begin{enumerate}
	\item Create a random source (set of coordinates).
	\item Create a random number with a range equal to the values of the map or mask.
	\item Compare the value of the map at those coordinates to the random number.
	\item If the value of the point on the map exceeds the random number, then the random source is included in the sample for further analysis. Otherwise a new random source is created and the process is repeated until the correct amount of sources is obtained.
\end{enumerate}

This will effectively generate random sources that are more likely to appear where the exposure is higher. This method was tested by generating a statistically significant number of random sources and checking that the resulting image was one proportional to the weighting mask well within normal statistical fluctuations.

\subsection{Contamination by Galactic sources}

Considering that i) the density of matter of the Milky Way is roughly correlated with the amount of interstellar dust, and by consequence so is the amount of stellar sources, and ii) the transparency of gamma-rays to interstellar dust, we used maps of dust IR emission ~\citep{1998ApJ...500..525S} as a weighting mask for the Metropolis-Hastings algorithm to generate random Galactic sources. 

The isotropically generated samples were contaminated by increasing the number of Galactically generated random sources (N) to observe how this affected the PMF of their dipolar and quadripolar moment. We considered all possible combinations for the number of GRBs in the different samples and took into account the \textit{Swift} exposure map for each generated source.

\section{Results}
\begin{figure}
	\includegraphics[width=0.5\textwidth]{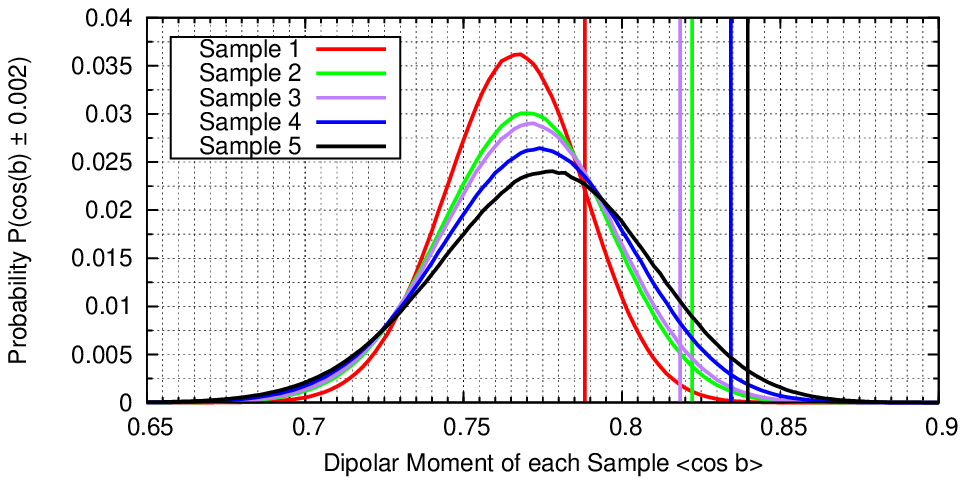}
	\includegraphics[width=0.5 \textwidth]{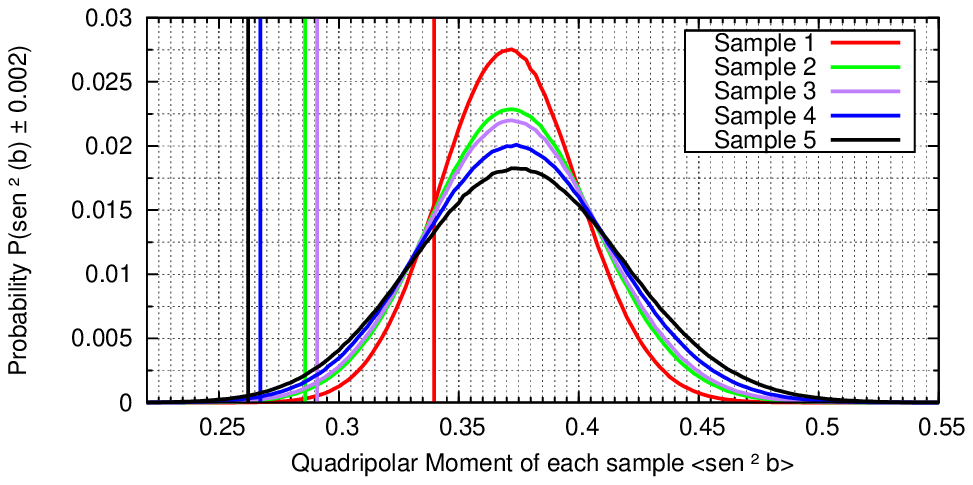}
	\caption{Dipolar (\textbf{top}) and quadripolar (\textbf{bottom}) moment PMFs for samples of isotropically generated sources weighted by 	\textit{Swift}'s exposure map for each sample size, and the observed values for each sample (vertical lines).}
	\label{isotropic}
\end{figure}

The Monte Carlo simulations of the isotropically generated random samples (weighted by the \textit{Swift} exposure map) showed that the dipolar and quadripolar moments from the real samples consistently deviated from the average, as is shown in Figure~\ref{isotropic}. 

Table \ref{Percentiles} lists the percentile of the population in which each sample is located. Considering that by definition, one standard deviation will be between the 15.87th and the 84.13th percentiles, two standard deviations between 2.28th and 97.72th and three deviations between 0.13th and 99.87th, we observe that with the exception of the first sample, all samples have dipolar and quadripolar moments located outside two standard deviations. 

\begin{figure}
	\includegraphics[width=0.5\textwidth]{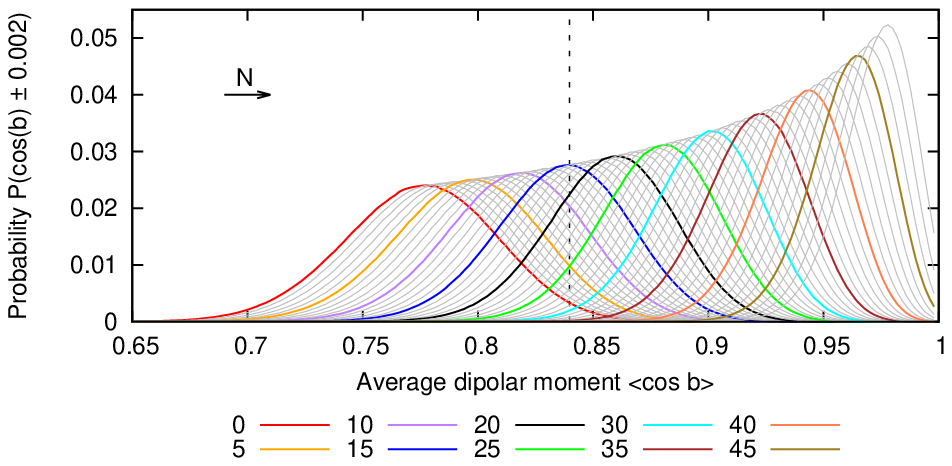}
	\includegraphics[width=0.5 \textwidth]{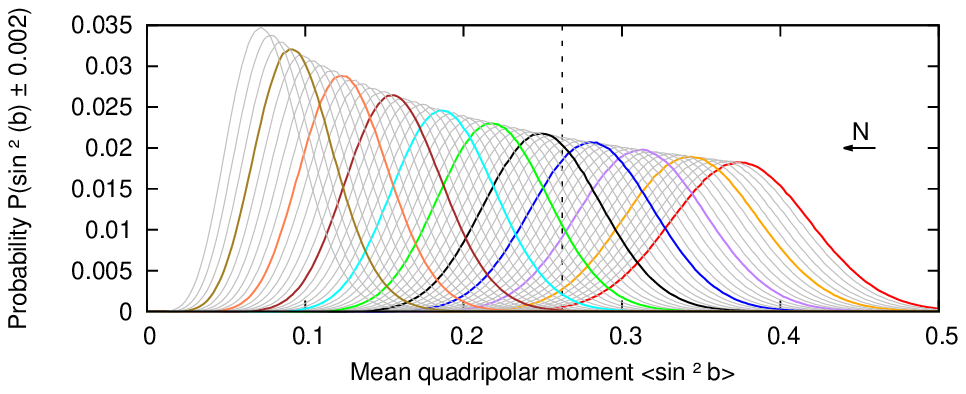}
	\caption{Example of dipolar (\textbf{top}) and quadripolar (\textbf{bottom}) moment probability distributions for samples of 49 randomly generated sources with an increasing number (\textbf{N}) of those sources being of Galactic origin, and the observed value for sample \#5 (dashed vertical line). Each line has a different amount of Galactic sources, starting with N=0 (red), and only every fifth line was colored for easier reading.}
	\label{49Gaussians}
\end{figure}

\begin{table}
\centering
	\begin{tabular}{|c|c|c|c|}
		\hline
		Sample & Dipolar  & Quadripolar \\ \hline
		\#1 & 78.24 & 13.26 \\ \hline
		\#2 & 97.91 & 0.59 \\ \hline
		\#3 & 96.72 & 1.03 \\ \hline
		\#4 & 98.52 & 0.44 \\ \hline
		\#5 & 97.83 & 0.39 \\ \hline
	\end{tabular}
	\caption{Percentile of the values observed for the dipolar and quadripolar moments of each sample.}
	\label{Percentiles}
\end{table}
The probability distribution of samples that contained both isotropically and Galactically generated sources allowed us to compare how the contamination by Galactic sources affected the likelihood of obtaining certain momentum values, for example see figure \ref{49Gaussians}. This technique is similar to the one used in the past for studying the degree of contamination by Galactic repeater gamma-ray sources present in two GRB catalogs \citep{1998A&A...336...57G}.

\begin{figure}
	\includegraphics[width=0.5\textwidth]{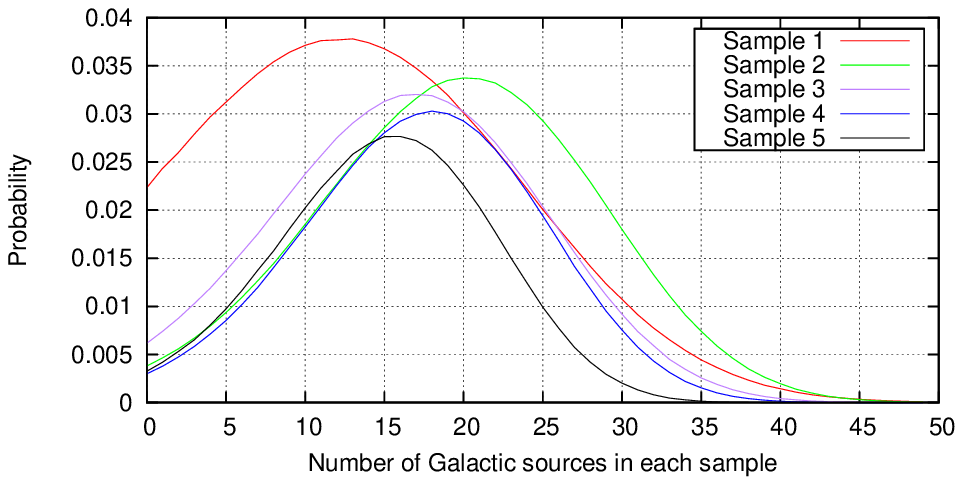}
	\includegraphics[width=0.5 \textwidth]{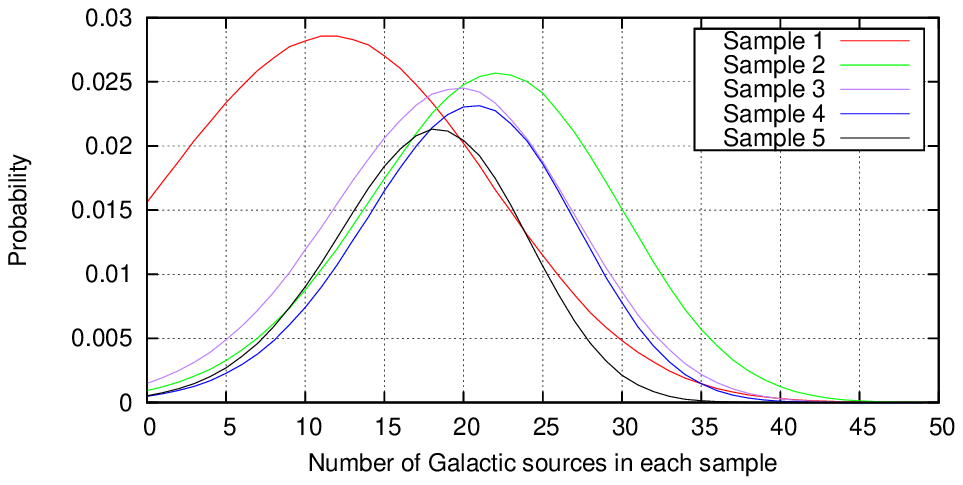}
	\caption{Relative probability of obtaining the dipolar (\textbf{top}) and quadripolar (\textbf{bottom}) moments $\pm 0.002$ measured for our samples. }
	\label{Probabilities}
\end{figure}

By observing the probability of the observed values in each one of the curves that resulted from the simulations, we determined the relative probability that each one of those combinations of isotropically and Galactically generated sources would yield the observed momentums. Figure \ref{Probabilities} shows the probability as a function of the amount of Galactic sources introduced in each sample.

\section{Conclusions}

With the exception of the first sample, all observed samples show dipolar and quadripolar moments outside two standard deviations from the mean of an isotropically generated distribution. Although this result is not conclusive, there is a high probability that the samples are not of a purely Extra-Galactic nature.

The probability of obtaining the dipolar and quadripolar moments that are measured in the samples is much higher when including a significant amount of Galactic sources than it is for only isotropically generated sources.

\begin{table}
\centering
	\begin{tabular}{|c|c|c|c|c|c|}
	\hline
	\textbf{Sample \#:} & 1 & 2 & 3 & 4 & 5 \\ \hline
	\textbf{Dipolar:} & 13 & 20 & 17 & 18 & 16 \\ \hline
	\textbf{Dipolar \%:} & 11.71\% & 25.97\% & 23.94\% & 30.51\% & 32.65\% \\ \hline
	\textbf{Quadripolar:} & 12 & 22 & 20 & 21 & 18 \\ \hline
	\textbf{Quad. \%:} & 10.81\% & 28.57\% & 28.17\% & 35.59\% & 36.73\% \\ \hline
	\end{tabular}
	\caption{Amount of Galactic sources that yield a higher probability to obtain the observed dipolar and quadripolar moments and the percentage of the sample size they imply.}
	\label{contamination}
\end{table}

As shown in Table \ref{contamination}, if we consider the amount of contamination that yields the highest probability of obtaining the observed values, the amount of Galactic sources that are probably contaminating the \emph{Swift} GRB catalog is between 16 and 22. This value represents approximately $3 \%$ of the catalog used for this study.

Sample 5 has been narrowed down so that it is likely that one out of every three is in fact a Galactic source, accordingly it is of great interest to study these sources in more detail to determine if there are other indications that they are not GRBs.

The high Galactic extinction discourages optical ground-based spectroscopy of most low Galactic latitude GRBs. We showed that a large part of those abandoned follow-ups could reveal a missing population of Galactic events. We encourage ground observers to follow-up those events, since it might lead to the discovery of unknown high-energy phenomena in our Galaxy.

\begin{acknowledgements}
We have made use of J. Greiner's GRB Table (http://www.mpe.mpg.de/~jcg/grbgen.html)
  
This research has been partially supported by the Spanish Ministry of Economy and Competitivity under the programmes AYA2011- 24780/ESP, AYA2009-14000-C03-01/ESP,  and AYA2012-39362-C02-02 and OTKA grant K077795.
\end{acknowledgements}

\bibliographystyle{aa} 
\bibliography{ref}

\onecolumn
This appendix specifies which bursts are included in each one of the five samples used in this study. To facilitate its reading each column was named in reference to the discriminating parameter for each subsample.
That is:

	\begin{itemize}
	\item \textbf{Sample 1:} All 111 FREDs detected by \textit{Swift} before March 2011.
	\item \textbf{Sample 2:} 77 FREDs without any measured redshift different from zero
	\item  \textbf{Sample 3:} 71 FREDs without any claimed high redshift criteria
	\item  \textbf{Sample 4:} 59 FREDs without any type of indirect redshift indication
	\item  \textbf{Sample 5:} 49 FREDs without any redshift indications or multiple peaks
	\end{itemize}

With the exception of the first column and where noted otherwise, each number refers to the GCN circular.

\begin{longtable}{|c|c|c|c|c|c|}
\hline
GRB\#	&F.R.E.D.	&Redshift	&"High-z"	&Any z	&Peaks\\\hline
\endhead
050315	&3094	&3101	&-	&-	&-\\\hline
050319	&3117	&3135	&-	&-	&-\\\hline
050406	&3183	&none	&none	&none	&none\\\hline
050421	&3305	&none	&none	&none	&none\\\hline
050603	&3512	&3520	&-	&-	&-\\\hline
050713B	&3600	&none	&none	&none	&none\\\hline
050717	&3633	&none	&none	&none	&3633\\\hline
050721	&3661	&none	&none	&none	&none\\\hline
050801	&3725	&none	&none	&none	&none\\\hline
050826	&3888	&none	&5982	&-	&-\\\hline
051016	&4102	&4391	&-	&-	&-\\\hline
051021B	&4126	&none	&none	&none	&none\\\hline
051111	&4260	&4255	&-	&-	&-\\\hline
051117A	&4289	&none	&none	&none	&none\\\hline
051227	&4401	&none	&none	&4399	&-\\\hline
060110	&4463	&none	&none	&none	&none\\\hline
060213	&4762	&none	&none	&4769	&-\\\hline
060403	&4945	&none	&none	&none	&none\\\hline
060515	&5141	&none	&none	&none	&none\\\hline
060522	&5153	&5155	&-	&-	&-\\\hline
060526	&5163	&5164	&-	&-	&-\\\hline
060604	&5212	&5218	&-	&-	&-\\\hline
060605	&5221	&5223	&-	&-	&-\\\hline
060607	&5242	&5237	&-	&-	&-\\\hline
060707	&5285	&5298	&-	&-	&-\\\hline
060719	&5349	&none	&none	&none	&none\\\hline
060904B	&5520	&none	&none	&5507	&-\\\hline
060912	&5558	&5565	&-	&-	&-\\\hline
060919	&5578	&none	&none	&none	&none\\\hline
060926	&5621	&5626	&-	&-	&-\\\hline
061102	&5777	&none	&none	&none	&none\\\hline
061110A	&5802	&5812	&-	&-	&-\\\hline
061202	&5887	&none	&none	&none	&none\\\hline
061210	&5905	&none	&none	&none	&none\\\hline
061222A	&5964	&none	&none	&Paper:\citep{2009AJ....138.1690P}	&-\\\hline
070208	&6081	&6080	&-	&-	&-\\\hline
070219	&6109	&none	&none	&none	&none\\\hline
070318	&6210	&6213	&-	&-	&-\\\hline
070326	&6653	&none	&none	&none	&none\\\hline
070330	&6237	&none	&6238	&-	&-\\\hline
070509	&6394	&none	&none	&none	&none\\\hline
070518	&6415	&none	&none	&none	&none\\\hline
070520B	&6438	&none	&none	&none	&none\\\hline
070531	&6475	&none	&none	&none	&none\\\hline
070612	&6509	&none	&none	&6510	&-\\\hline
070808	&6718	&none	&none	&6720	&-\\\hline
070810B	&6753	&none	&none	&6756	&-\\\hline
070917	&6791	&none	&none	&6799	&-\\\hline
071010B	&6871	&6884	&-	&-	&-\\\hline
071028	&7013	&none	&none	&none	&none\\\hline
071028B	&7019	&none	&none	&Paper:\citep{2011AA...529A.110C}	&-\\\hline
071112C	&7081	&7070	&-	&-	&-\\\hline
071117	&7098	&none	&none	&7108	&-\\\hline
080303	&7351	&none	&none	&none	&none\\\hline
080307	&7362	&none	&7362	&-	&-\\\hline
080319C	&7442	&7468	&-	&-	&-\\\hline
080320	&7473	&none	&none	&7474	&-\\\hline
080325	&7531	&none	&none	&none	&none\\\hline
080413B	&7606	&7598	&-	&-	&-\\\hline
080430	&7647	&7647	&-	&-	&-\\\hline
080503	&7673	&none	&none	&none	&7673\\\hline
080517	&7748	&none	&7748	&-	&-\\\hline
080523	&7772	&none	&7772	&-	&-\\\hline
080613B	&7876	&none	&none	&none	&7873\\\hline
080701A	&7913	&none	&none	&none	&none\\\hline
080702B	&7924	&none	&none	&none	&none\\\hline
080707	&7952	&7948	&-	&-	&-\\\hline
080710	&7969	&7962	&-	&-	&-\\\hline
080714	&7979	&none	&none	&none	&none\\\hline
080727B	&8030	&none	&8033	&-	&-\\\hline
080805	&8059	&8060	&-	&-	&-\\\hline
080903	&8176	&none	&none	&none	&none\\\hline
080915B	&8234	&none	&none	&none	&none\\\hline
080916A	&8243	&8254	&-	&-	&-\\\hline
081104	&8479	&none	&none	&none	&none\\\hline
081121	&8537	&8542	&-	&-	&-\\\hline
090129	&8861	&none	&none	&none	&none\\\hline
090401B	&9068	&none	&none	&none	&9066\\\hline
090518	&9393	&none	&none	&none	&none\\\hline
090520	&9417	&none	&none	&none	&none\\\hline
090530	&9443	&none	&none	&none	&9443\\\hline
090726	&9706	&9712	&-	&-	&-\\\hline
090727	&9724	&none	&none	&none	&9724\\\hline
090904A	&9888	&none	&none	&none	&none\\\hline
091018	&10034	&10038	&-	&-	&-\\\hline
091024	&10072	&10065	&-	&-	&-\\\hline
091026	&10081	&none	&none	&none	&none\\\hline
091208A	&10253	&none	&none	&none	&none\\\hline
091208B	&10265	&10263	&-	&-	&-\\\hline
100111A	&10317	&none	&none	&none	&none\\\hline
100115A	&10325	&none	&none	&none	&none\\\hline
100213B	&10412	&10422	&-	&-	&-\\\hline
100316A	&10501	&none	&none	&none	&none\\\hline
100316B	&10500	&10495	&-	&-	&-\\\hline
100423A	&10651	&none	&none	&none	&10658\\\hline
100514A	&10761	&none	&none	&none	&none\\\hline
100702A	&10926	&none	&none	&none	&none\\\hline
100704A	&10929	&none	&none	&10940	&-\\\hline
100727A	&11001	&none	&none	&none	&none\\\hline
100802A	&11031	&none	&none	&none	&none\\\hline
100814A	&11094	&11089	&-	&-	&-\\\hline
100823A	&11135	&none	&none	&none	&none\\\hline
100917A	&11289	&none	&none	&none	&none\\\hline
101008A	&11318	&none	&none	&none	&11318\\\hline
101011A	&11332	&none	&none	&none	&11331\\\hline
101017A	&11345	&none	&none	&none	&11345\\\hline
101023A	&11363	&none	&none	&none	&none\\\hline
101114A	&11405	&none	&none	&none	&none\\\hline
101213A	&11448	&11457	&-	&-	&-\\\hline
110102A	&11509	&none	&none	&none	&none\\\hline
110223A	&11764	&none	&none	&none	&none\\\hline
Total:	&111	&77	&71	&59	&49\\\hline
\caption{GRBs of the samples used for this study and the GCN used to discriminate each burst.}
\label{samplestable}
\end{longtable}
\twocolumn

\end{document}